# Structural and dielectric characterization of Sm$_2$MgMnO$_6$

Moumin Rudra*, Ritwik Maity, T. P. Sinha

Department of Physics, Bose Institute, 93/1, Acharya Prafulla Chandra Road, Rajabazar, Kolkata – 700009, India

The polycrystalline Sm2MgMnO6 (SMMO) was synthesized at 1173K by means of sol-gel technique. Rietveld refinement of X-ray diffraction (XRD) pattern confirmed the formation of a single phase monoclinic structure with space group P2$1/n$. The band gap achieved from UV-vis spectra shows the semiconducting nature of the material. To observe the effect of grains and grain-boundaries in the conduction process and dielectric relaxation measurements are carried out on SMMO sample at different frequencies between 313 K and 673 K. An electrical equivalent circuit consisting of the resistance and constant phase element is used to clarify the impedance data.

## 1. Introduction

Previous years, there is a great activity in developing new multifunctional double perovskite oxides (DPOs), which are very charming for their potential application in solid oxide fuel cells, thin film substrate for superconductors, electrolytes also in spintronics and magnetoelectric devices [1-5]. The alternating current impedance spectroscopy (ACIS) is very acceptable and powerful diagnostic tool for investigating the electrical properties of dielectric materials [6-8]. It helps to investigate the capacitance and conductance over a wide range of frequency range at various temperatures. This method describes the electrical processes arise in a system by applying an ac signal as an input perturbation. Hence, these days' ACIS has become very popular method in materials research and development. The dielectric studies of DPOs have been the focus of researchers in recent years due to their scientific as well as technological importance and some considerable works have been already reported [9-11]. Though the electrical properties of various DPOs have been studied, no attempt has been made to study the DPO Sm$_2$MgMnO$_6$. In this paper we, therefore, report the dielectric relaxation behaviour of a new DPO Sm$_2$MgMnO$_6$ (SMMO) compound prepared by sol-gel technique in a wide range of frequency (42 Hz – 5 MHz) and temperature (313 K–673 K) using impedance spectroscopy.

## 2. Experimental

SMMO was synthesized by the sol–gel citrate method. At first, stoichiometric quantities of the reagent grade Sm(NO$_3$)$_3$, 6H$_2$O, Mg(NO$_3$)$_2$, 9H$_2$O and Mn(NO$_3$)$_2$, 9H$_2$O were taken and dissolved separately in de-ionized water by stirring using a magnetic stirrer. citric acid (CA) and ethylene glycol (EG) were added to this solution drop wise according to the molar ratio of {Sm$^{3+}$}:{CA}:{EG} = 1:1:4 to form a polymeric-metal cation network. The solution was stirred at 303 K using a magnetic stirrer for 6 h to get a homogeneous solution which was dried at 393 K to obtain the gel precursor. After combustion of the gel, a fluffy powder of the material was collected. the powder was calcined at 1173 K in the air for 12 h and cooled down to room temperature (RT~300K) at a cooling rate of 1 K/min. the disc of thickness 0.9 mm and diameter 8 mm was prepared by the calcined sample using polyvinyl alcohol (PVA) as the binder. finally, the disc was sintered at 1273 K and cooled down to RT by cooling at the rate of 1 K/min. the crystal-structure of SMMO was studied using a X-ray powder diffractometer (Rigaku Miniflex II) having Cu-K$_\alpha$ radiation in the 2θ range of 10–80° by scanning at 0.02° per step at RT. the refinement of the X-ray diffraction (XRD) pattern was performed by the Rietveld method with the full-prof program [12]. In the refinement process of the XRD profile, the background was fitted with the 6-coefficients polynomial function, and the peak shapes were described by the pseudo-voigt profiles. During the refinement, the scale factor, lattice parameters, positional coordinates (x, y, z) and thermal parameter (B$_{iso}$) were varied whereas the occupancy parameters of all the ions were kept fixed.

In order to study the microstructure of SMMO, the pellet has been fractured and is kept on a stub with gold coating on the surfaces of the sample. The microstructural images are taken using FEI quanta 200 scanning electron microscope to determine the grain size distribution, surface morphology. Standard UV–visible absorption spectrum of the sample was obtained in the range of 200–900 nm using a Shimadzu UV–visible spectrometer. To study the electrical properties, both the flat surfaces of the sintered pellet were electroded with the thin gold pellet to performing the experiment. The impedance (Z), capacitance (C$_S$), conductance (G) and phase angle (φ) were measured using an LCR meter (Hioki) in the frequency range from 42 Hz to 5 MHz at the oscillation voltage of 1.0 V. the measurements were performed over the temperature range from 313 K to 673 K using an in-built cooling–heating system. The temperature was controlled by a Eurotherm 818p programmable temperature controller connected with the oven. Each measured temperature was kept constant with an accuracy of ±1 K. The complex dielectric constant ε* (=1/iωC$_0$Z*) was obtained from the temperature dependence of the real (Z') and imaginary (Z'') parts of the complex impedance Z* (=Z' - iZ''), where ω is the angular frequency (ω = 2πν) and i = √(-1). C$_0$ = ε$_0$A/d is the empty cell capacitance, where A is the sample area and d is the sample thickness. The ac electrical conductivity σ (= Gd/A) was calculated from the conductance.

## 3. Results and Discussion
### A. X-ray diffraction

Fig. 1 shows the XRD pattern of SMMO where the symbol represents the best fit to the diffraction pattern obtained by Rietveld refinement [13]. The vertical bar symbols denote the Bragg-positions and the solid curve at the bottom represents the difference between the experimental and the calculated patterns. It has inferred that SMMO may crystallize in a perovskite structure with monoclinic $P21/n$ space group. The presence of the super-lattice diffraction peak (101) at $2\theta = 20°$ in the XRD pattern indicates in phase tilting of the octahedra with B-site cation ordering. Hence the centrosymmetric space group $P21/n$, which permits B-sites ordering is adopted here to refine the crystal structure of SMMO.

*moumin.nitp@gmail.com

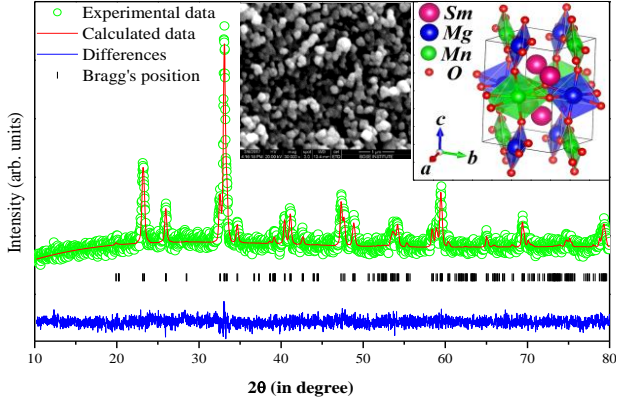

Fig.1 – Rietveld refinement plot of SMMO at room temperature. Inset (right) shows the schematic presentation of the SMMO monoclinic unit cell. The scanning electron micrograph of the sample as shown in inset (middle).

Table 1 – Various structural parameters extracted from Rietveld-refinement of XRD data for SMMO.

| Space group = $P21/n$ (Monoclinic) | | | |
|---|---|---|---|
| Lattice parameters: $a$ = 5.3667 Å; $b$ = 5.5012 Å; $c$ = 7.6299 Å; $\beta$ = 90.0833° | | | |
| Refinement parameters: $R_p$ = 5.70; $R_{wp}$ = 7.14; $R_{exp}$ = 6.63 and $\chi^2$ = 1.16 | | | |
| Atom | Wyckoff site | x | y | z |
| Sm | 4e | 0.4966 | 0.5481 | 0.2523 |
| Mg | 2c | 0 | 0.5000 | 0 |
| Mn | 2d | 0.5000 | 0 | 0 |
| O1 | 4e | 0.7487 | 0.3076 | 0.0135 |
| O2 | 4e | 0.1975 | 0.7917 | 0.0955 |
| O3 | 4e | 0.5999 | 0.0021 | 0.2612 |
| Bond lengths (Å) | Mg-O1(×2): 1.71(10) | Mn-O1(×2): 2.16(9) | |
| | Mg-O2(×2): 2.06(9) | Mn-O2(×2): 2.12(9) | |
| | Mg-O3(×2): 1.90(13) | Mn-O3(×2): 2.06(13) | |
| Bond angles (°) | Mg-O1-Mn: 166(4) | | |
| | Mg-O2-Mn: 134(3) | | |
| | Mg-O3-Mn: 149(5) | | |

The refined lattice parameters are found to be $a$ = 5.3667 Å; $b$ = 5.5012 Å; $c$ = 7.6299 Å; $\beta$ = 90.0833°. A schematic presentation of the SMMO unit cell is shown in the inset of Fig. 1 with the distribution of ions at crystallographic positions $4e$ for $Sm^{3+}$ ions, $2c$ for $Mg^{2+}$ ions, $2d$ for $Mn^{4+}$ ions, and $4e$ for $O^{2-}$ ions as given in Table 1. Each of $Mg^{2+}$ and $Mn^{4+}$ ions is surrounded by six $O^{2-}$ ions, constituting $MgO_6$ and $MnO_6$ octahedra respectively which are arranged alternatively. The final structure parameters along with the bond lengths and the bond angles associated with $MgO_6$ and $MnO_6$ octahedra are listed in Table I. It is well known that for an unmixed DPO of $A_2B'B''O_6$ type, the degree of distortion is determined by the tolerance factor, $t_f = (r_A + r_O)/\sqrt{2}(\langle r_B \rangle + r_O)$. Where $r_A$ and $r_O$ denote the ionic radii of A and O-ions, respectively and $\langle r_B \rangle$ denote the average ionic radius for the ions on the B site. Usually a cubic perovskite structure is obtained for $t_f$ close to unity, whereas it can be lower symmetrical structure (orthorhombic, monoclinic) for lower value of tolerance factor. The theoretical value of $t_f$ for SMMO is 0.86 which supports the lower symmetrical structure of the sample.

### B. Microstructural and optical studies

The SEM micrograph of the surface morphology of the SMMO ceramic is shown in inset (middle) of Fig. 1. The particles are nearly spherical and unequal in shape. The average grain size range 0.275-0.400 µm. The density of the sample is found to be 7.019 gm/cm$^3$.

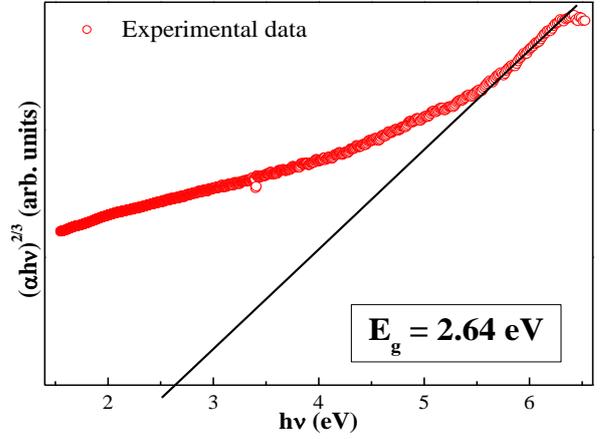

Fig.2 – *Tauc plot for SMMO sample.*

The band gap of SMMO has been estimated from the absorbance coefficient (α) data as a function of wavelength using Tauc relation [14].

$$\alpha \propto \frac{(h\nu - E_g)^n}{h\nu} \qquad (1)$$

where, $h\nu$ is the incident photon energy and $E_g$ is the optical band gap. The exponent n is a dimensionless parameter having value of 3/2 for direct forbidden transitions. The $(\alpha h\nu)^{2/3}$ vs. $h\nu$ plot shows the linear nature near the absorption edge for SMMO as shown in the Fig. 2. The extrapolation of the linear part of this curve near the absorption edge to $(\alpha h\nu)^{2/3} = 0$ axis gives the direct band gap energy of 2.64 eV.

### C. Dielectric relaxation and ac conductivity

The angular frequency dependence of dielectric constant (ε') and loss tangent ($tan\delta$) of SMMO at various temperatures is shown in Fig.3. The presence of two well resolved relaxation peaks in Fig. 3(b) confirms the existence of at least two types of relaxation process in SMMO. In the higher frequency range ($>10^4$ Hz) the relaxation process corresponds to the grain effect and in the lower frequency side ($<10^4$ Hz) the grain boundary contribution is more dominant. To explain the relaxation phenomenon in each frequency region, we have considered the Debye model. According to this model, below the relaxation frequency of each relaxation process all the dipoles follow the applied field and fully contribute to the relaxation process. With the increase of the applied field frequency, the dipoles begin to lag behind the applied field and at the relaxation frequency a sudden drop of the $\varepsilon'$ is evident. For further increase of the applied frequency, most of the dipoles do not respond and $\varepsilon'$ becomes nearly independent of the frequency. But this feature is not clearly observed in Fig. 3(a), may be due to the overlapping of the frequency range associated with more than one relaxation process.

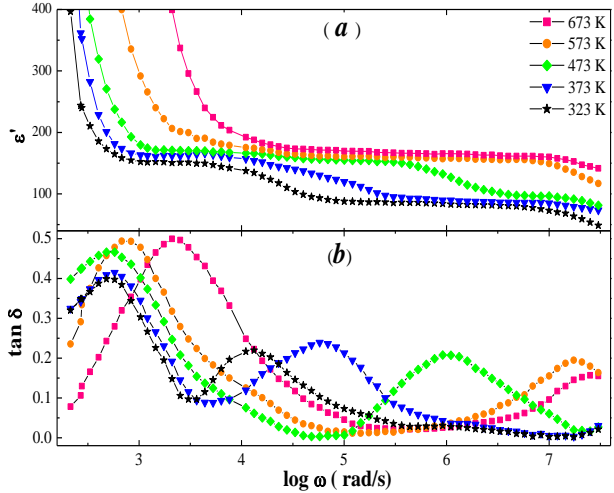

Fig. 3 – Frequency (angular) dependence of $\varepsilon'$ (a) and $tan\delta$ (b) at various temperatures for SMMO.

It is observed from Fig. 3 that the dispersions in $\varepsilon'$ and the corresponding relaxation peaks in $tan\delta$ move towards the higher frequencies with the increase of temperature which suggests the thermally activated nature of the relaxation process. Since the peak in the $tan\delta$ depends on the mobility and the temperature, the mobility of the thermally activated charge carriers increases with the increase of the temperature, and they start to relax at the higher frequency thereby shifting the loss peak towards the higher frequency side. It is observed from Fig. 3(b) that the value of $tan\delta$ in the grain region is smallest (<0.3) with respect to the grain-boundary (≥0.3) regions. Since the increase in the value of $\varepsilon'$ with the increase of the temperature is more pronounced in the lower frequencies, the observed high value of $\varepsilon'$ and $tan\delta$ at the lower frequency side can be attributed to the presence of the electrode–semiconductor interface, which results in the Maxwell–Wagner type polarization [15]. The Maxwell–Wagner type polarization may be originated due to the presence of heterogeneous components in the material, which have different interfaces with different conductivity. The surface charge accumulation is taken place when the current is passed through these interfaces which gives a Debye like relaxation under the application of an ac voltage. At the electrode–semiconductor contact, a high capacitance is formed due to the Schottky-type barrier layer which may be originated due to different work functions of the charge carriers at the electrode and in the materials [16]. This results in the high dielectric constant at the lower frequency and at the higher temperature in SMMO. The transport mechanism in SMMO can be assessed by the hopping of charge carriers, and the frequency dependent $tan\delta$ is a requisite part of the charge carriers hopping transport process.

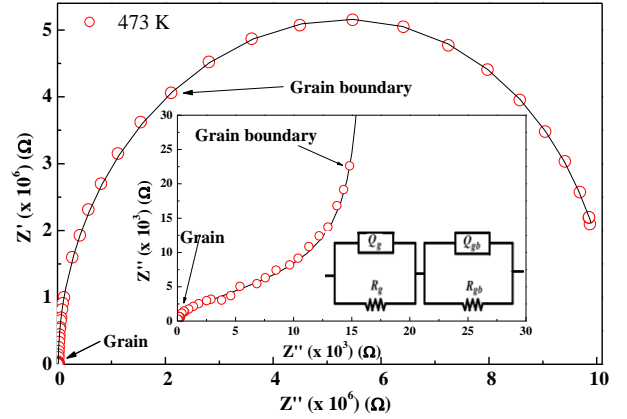

Fig.4 – The complex impedance plane plots between Z" and Z' at the temperature 473 K. Lower inset shows the two well resolved semicircular arcs at high frequency corresponding to grain and grain boundaries. Right inset shows the equivalent circuit model used for fitting non-ideal (Cole–Cole) behaviour.

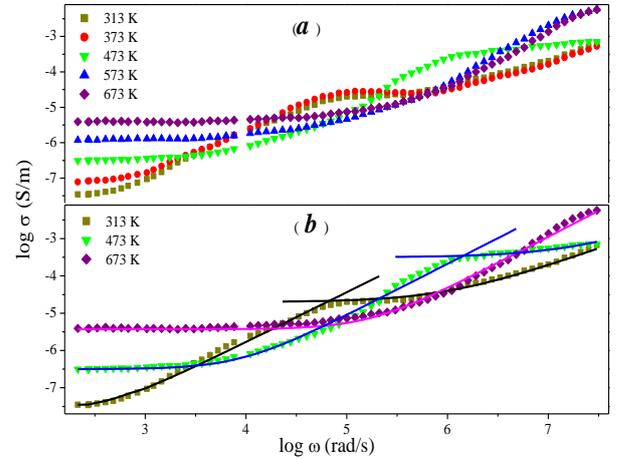

Fig.5 – Frequency (angular) dependence of the ac conductivity (σ) at various temperatures. The solid lines are the fitting of the experimental data with power law.

In order to understand the origin of the different relaxation processes in SMMO, we have studied its complex impedance plane plot (Z-plot) as shown in Fig. 4. The presence of two semicircular arcs in Fig. 4 confirms that two different types of relaxation process are involved in the charge transport of SMMO. The inset of Fig. 4 shows the high frequency data for the clarity of the grain and grain-boundary regions. Usually the Z-plot is fitted with an electrical equivalent circuit consisting of two parallel resistance–capacitance (RC) circuits connected in series. One parallel branch is associated with the grain effect and the other one represent the grain-boundary effects. Due to the non-ideal behaviour of the capacitance, sometimes both grain and grain-boundary contributions, though small, are present in the same frequency range, which may give rise to the depressed arcs or even only a spike-like nature in the low frequency region with a small arc in the high-frequency region of the Z-plot. For such cases, the capacitance term in the RC-equivalent circuit is replaced by a constant phase element (CPE). The capacitance of CPE can be expressed as $C_{CPE} = Q^{1/k}R^{(1-k)/k}$, where $k$ estimates the non-ideal behaviour and $Q$ is the CPE component. The value of $k$ is zero for the ideal resistance and 1 for the ideal capacitance. The solid lines in Fig. 4 represent the fitting to the electrical equivalent circuit and the fitted parameters are found to be to be 3000 Ω, 10320 kΩ, 1.0nF, 0.155nF, 0.85, and 0.9 for $R_g$, $R_{gb}$, $C_g$, $C_{gb}$, $k_g$ and $k_{gb}$ respectively.

Table 2 – The various fitted parameters of conductivity spectra in SMMO.

| Temperature (K) | $A_1$ (×10$^{-12}$) | $n_1$ | $A_2$ (×10$^{-11}$) | $n_2$ |
|---|---|---|---|---|
| 313 | 7.299 | 1.40 | 3.981 | 0.95 |
| 473 | 1.104 | 1.38 | 9.338 | 0.90 |
| 673 | 0.177 | 1.34 | - | - |

The frequency dependent log–log plots of the ac conductivity of SMMO are shown in Fig. 5(a) at a various temperatures. A strong frequency dependent and a very week temperature dependent effect are observed for the ac conductivity in SMMO. Due to the effect of grain and grain-boundary, one gets two plateaus and two dispersion regions in the frequency dependent ac conductivity plots as shown in Fig. 5(b) [17]. The low frequency plateau represents the grain boundary contribution to the total conductivity. The grain-boundary contribution relaxes in the dispersion region after this plateau. The highest frequency plateau represents the contribution of grains to the total conductivity. The dispersion region followed by this plateau represents the frequency dependence of the bulk conductivity. At the higher temperature, the low frequency plateau becomes nearly independent of frequency and its value at $\nu \to 0$ gives the value of dc conductivity at that temperature. This is due to the fact that the intrinsic conductivity is dominant in the low frequency region at the higher temperatures. The increase of the conductivity with the increase of the frequency at a particular temperature indicates that the ac conductivity follows the power law. Since two plateaus are present in the ac conductivity we have used the power law equation having two frequency dependent parts defined as follows:

$$\sigma(\omega) = \sigma_0 + A_1\omega^{n_1} + A_2\omega^{n_2} \qquad (2)$$

where $\sigma_0$ is the frequency independent conductivity, the coefficients $A_1$, $A_2$ and $n_1$, $n_2$ are the temperature and material dependent parameters. The experimental conductivity data are fitted by eq$^n$ (2) as shown by the solid lines in Fig. 5(b) for 313, 473 and 673 K. The values of the fitted parameters are given in Table 2. The temperature dependence of $n_1$ and $n_2$ gives the information to specify the suitable mechanism involved for the ac conductivity. It is observed that the values of $n_1$ and $n_2$ decrease with an increase in the temperature which can be explained due to the hopping between uncorrelated pairs of hopping centres, i.e., the short range translational hopping in the low and mid-frequency regions.

### 4. Conclusions

The DPO Sm$_2$MgMnO$_6$ (SMMO) has been synthesized by the sol–gel technique. The Rietveld refinement of the XRD profile at the room temperature shows the monoclinic $P2_1/n$ crystal symmetry of the system. The crystal structure of SMMO differs from its two end materials SmMgO$_3$ and SmMnO$_3$. The effect of grain and grain-boundary in the relaxation process is explained from the frequency dependent dielectric constant and loss tangent. An electrical equivalent circuit consisting of resistance and constant phase element is used to explain the complex impedance plane plot. The frequency dependent ac conductivity spectra follow the power law equation having two frequency dependent parts for the effect of grain and grain-boundary.

### 5. Acknowledgement

Moumin Rudra acknowledges the financial support provided by the UGC New Delhi in the form of JRF. Ritwik Maity thanks to Department of Science and Technology of India for providing INSPIRE fellowship in the form of JRF.

M. Rudra, R. Maity, T. P. Sinha